\documentclass[final,5p,twocolumn,times,authoryear]{elsarticle}

\usepackage{tabularx}
\usepackage{ragged2e}
\usepackage{comment}
\usepackage{booktabs}
\usepackage{amssymb}
\usepackage{amsmath}
\usepackage{tipa}
\usepackage{xurl}

\journal{Speech Communication}

\usepackage{xcolor}

\usepackage{lineno}
\begin{document}

\begin{frontmatter}

\title{The Voiceprint Fallacy: Why Voices Are Not Unique Biometric Imprints}

\author[inst1]{Tianle Yang}
\author[inst2,inst3]{Cuiling Zhang\corref{cor1}}
\author[inst4]{Chengzhe Sun}
\author[inst4]{Siwei Lyu}
\author[inst5]{Phil Rose}

\cortext[cor1]{Corresponding author: cuiling-zhang@forensic-voice-comparison.net}

\affiliation[inst1]{
organization={University at Buffalo},
addressline={Department of Linguistics},
city={Buffalo},
state={NY},
country={United States}
}

\affiliation[inst2]{
organization={Southwest University of Political Science and Law},
addressline={Institute of Intelligent Justice},
city={Chongqing},
country={China}
}

\affiliation[inst3]{
organization={Southwest University of Political Science and Law},
addressline={School of Criminal Investigation},
city={Chongqing},
country={China}
}

\affiliation[inst4]{
organization={University at Buffalo},
addressline={Department of Computer Science and Engineering},
city={Buffalo},
state={NY},
country={United States}
}

\affiliation[inst5]{
organization={Australian National University},
addressline={Emeritus Faculty},
city={Canberra},
state={ACT},
country={Australia}
}

\begin{abstract}
In recent years, the term voiceprint has regained attention, particularly in technological applications and policy-making contexts, often carrying the assumption that a person's voice constitutes a stable and unique biometric trace analogous to a fingerprint. Yet this conception has been repeatedly criticized and rejected by forensic voice experts throughout the decades since its introduction. Although voices undoubtedly contain speaker-related information, this simplified conception obscures the highly dynamic and context-dependent nature of speech. This article revisits the voiceprint fallacy and reconsiders what can count as evidence of speaker identity by reviewing the historical development of voiceprint identification, evidence on human voice variability, developments in forensic voice comparison, research on human and automatic speaker recognition, and the recent challenge posed by deepfake speech to speaker identity. We point out that the voiceprint metaphor and its underlying implications are scientifically misleading because they transform a probabilistic source of speaker information into an imagined stable object of identity. We argue that speaker identity assessment does not require, and current evidence does not support, the existence of a stable and individually unique voiceprint. For speaker recognition and voice biometrics, this distinction motivates interpreting learned speaker representations with respect to the conditions under which they are trained and evaluated, and explicitly assessing their robustness to relevant sources of within-speaker variability, domain mismatch, and synthetic manipulation.
\end{abstract}

\begin{comment}
\begin{highlights}
\item Shows voice variability across physiological, linguistic, and recording conditions.
\item Explains speaker individuality through within- and between-speaker variability.
\item Advocates validated probabilistic approaches to speaker identity assessment.
\item Shows that similarity does not by itself establish speaker identity.
\end{highlights}
\end{comment}

\begin{keyword}
voiceprint \sep voice biometrics \sep speaker recognition \sep speaker identity \sep vocal uniqueness
\end{keyword}

\end{frontmatter}

\section{Introduction}
The human voice conveys multiple kinds of information, including cues to a speaker's identity. In forensic and technological settings, speech has long been used for speaker comparison and authentication \citep{hansen2015speaker}. The modern forensic use of the term voiceprint is commonly traced to Kersta's article "Voiceprint Identification" \citep{kersta1962voiceprint}, which promoted the use of spectrographic representations of speech for speaker comparison. The visual appearance of spectrograms encouraged an analogy with fingerprints: if fingerprints are visible traces of identity, perhaps voices could be treated like a visible imprint as well. The terms "voiceprint identification" and "voiceprint" subsequently came to carry two related meanings. In its original methodological sense, it referred to the forensic practice of visually comparing speech spectrograms to determine whether two recordings were produced by the same speaker. In a broader derived sense, it came to suggest that each person possesses a stable and uniquely identifiable vocal biometric trace, like a visible imprint.

Both interpretations, however, have long been criticized by forensic specialists. \cite{rose2002forensic}, for example, argued that the original voiceprinting method promoted by Kersta is inappropriate because it is an "unquantified gestalt comparison of the way voices sound and look on spectrograms". More recently, the OSAC Standard Guide to the Forensic Speaker Recognition Landscape stated that beliefs concerning the scientific validity of voiceprinting have been rejected and discredited by the wider speaker recognition community \citep{osac2024}. Nevertheless, the term and its underlying metaphor continue to appear in various authoritative documents, including legal and policy definitions of biometric identity. For example, U.S. federal law defines a “means of identification” as including “unique biometric data,” such as a “voice print,” under 18 U.S.C. \S~1028 \citep{usc1028}. Similarly, regulations issued by the U.S. Department of Education identify voiceprints as examples of biometric records that may be used for the automated recognition of an individual, under 34 C.F.R. \S~99.3 \citep{cfr99}. Beyond such policy definitions, the term also appears in commercial voice systems \citep{microsoft2021,aws2021voiceid}, technical research on voice privacy, voice identity, and synthetic speech \citep{deng2023catch,zhang2023study,babu2024programmable,xue2026profiling}, and institutional IRB materials \citep{utah2025hipaa, ubirb2026hrp503, stanford2026confidentiality}.

\begin{table*}[t!]
\centering
\caption{Illustrative examples of vocal-uniqueness language in contemporary technical literature. Section A lists explicit person-level claims of vocal uniqueness. Section B lists weaker operational or representational formulations in which speaker-specific patterns, signatures, characteristics, embeddings, or voiceprints are described as unique. Inclusion in the table does not imply that the corresponding method requires or demonstrates absolute population-level vocal uniqueness, and the examples are not intended to quantify the prevalence of such claims in the literature.}
\label{tab:voice_uniqueness}

\footnotesize
\setlength{\tabcolsep}{5pt}
\renewcommand{\arraystretch}{1.12}

\begin{tabularx}{\textwidth}{
@{}
>{\RaggedRight\arraybackslash}p{0.065\textwidth}
>{\RaggedRight\arraybackslash}p{0.405\textwidth}
>{\RaggedRight\arraybackslash}X
@{}
}
\toprule
\textbf{Year} &
\textbf{Study and context} &
\textbf{Uniqueness formulation} \\
\midrule

\multicolumn{3}{@{}l}{\textbf{A. Explicit person-level claims of vocal uniqueness}}\\
\addlinespace[2pt]

2018 &
\textit{Redesign of Gaussian Mixture Model for Efficient and Privacy-Preserving Speaker Recognition}
(Speaker recognition / privacy) \citep{rahulamathavan2018redesign} &
Explicitly states that ``the voice is unique for individuals'' and treats voice features and enrolled voice templates as individual-specific biometric information. \\

\addlinespace[3pt]

2019 &
\textit{An Ultrathin Conformable Vibration-Responsive Electronic Skin for Quantitative Vocal Recognition}
(Voice biometrics / sensing) \citep{lee2019ultrathin} &
States that ``each entity has a unique voice pattern'' and uses this premise to motivate voice authentication. \\

\addlinespace[3pt]

2020 &
\textit{Replay Spoofing Countermeasure Using Autoencoder and Siamese Networks on ASVspoof 2019 Challenge}
(Automatic speaker verification) \citep{adiban2020replay} &
States that each individual has a ``unique voice pattern'' that is identifiable as a signature. \\

\addlinespace[3pt]

2021 &
\textit{A Review on Speaker Recognition: Technology and Challenges}
(Speaker-recognition review) \citep{hanifa2021review} &
Explicitly states that every individual's voice is unique and relates this claim to anatomical differences in the speech-production system. \\

\addlinespace[3pt]

2024 &
\textit{Milestones in Speaker Recognition}
(Speaker-recognition review) \citep{sharma2024milestones} &
Explicitly states that, like an individual's DNA, an individual's voice is unique, and subsequently describes speech as a biometric marker or ``voice-print.'' \\

\addlinespace[3pt]

2025 &
\textit{Human Voice is Unique}
(vocal uniqueness) \citep{singh2025human} &
Explicitly examines the hypothesis that human voice is unique and proposes a statistical framework for estimating the probability that two individuals could have the same voice. \\

\addlinespace[6pt]
\multicolumn{3}{@{}l}{\textbf{B. Operational or representational uniqueness language}}\\
\addlinespace[2pt]

2017 &
\textit{Evaluation of a Speaker Identification System with and without Fusion Using Three Databases in the Presence of Noise and Handset Effects}
(Speaker identification) \citep{alkaltakchi2017speaker} &
Describes speaker identification as identifying speakers on the basis of their ``unique voice pattern.'' \\

\addlinespace[3pt]

2021 &
\textit{Spoofing Speaker Verification System by Adversarial Examples Leveraging the Generalized Speaker Difference}
(Speaker verification / security) \citep{luo2021spoofing} &
Describes an utterance as containing a ``unique biometrics feature called voiceprint,'' which is used for speaker verification. \\

\addlinespace[3pt]

2022 &
\textit{Selective Listening by Synchronizing Speech with Lips}
(Speaker extraction) \citep{pan2022selective} &
States that each speaker has a ``unique voice signature'' and that this signature can be characterized by a fixed-dimensional speaker embedding. \\

\addlinespace[3pt]

2022 &
\textit{USEV: Universal Speaker Extraction with Visual Cue}
(Speaker extraction) \citep{pan2022usev} &
States that each speaker has a ``unique voice signature'' that can be characterized by speaker embeddings such as i-vectors, x-vectors, and d-vectors. \\

\addlinespace[3pt]

2022 &
\textit{RACP: A Network with Attention Corrected Prototype for Few-Shot Speaker Recognition Using Indefinite Distance Metric}
(Few-shot speaker recognition) \citep{wang2022racp} &
Defines speaker recognition as identifying a speaker from the ``unique voiceprint of utterances.'' \\

\addlinespace[3pt]

2023 &
\textit{ECAPA-Based Speaker Verification of Virtual Assistants: A Transfer Learning Approach}
(Speaker verification) \citep{rani2023ecapa} &
Characterizes speaker-verification technology as identifying individuals through their ``unique voice characteristics.'' \\

\addlinespace[3pt]

2023 &
\textit{Multi-Task Deep Cross-Attention Networks for Far-Field Speaker Verification and Keyword Spotting}
(Speaker verification) \citep{liang2023multitask} &
Describes speaker verification as relying on ``unique voice characteristics'' and characterizes a speaker embedding as representing the unique characteristics of the speaker's voice. \\

\addlinespace[3pt]

2023 &
\textit{Enhanced-Deep-Residual-Shrinkage-Network-Based Voiceprint Recognition in the Electric Industry}
(Voiceprint recognition) \citep{zhang2023enhanced} &
Motivates voiceprint recognition in part through the stated ``uniqueness of the voiceprint features,'' which the authors relate to speaker-specific vocal physiology. \\

\addlinespace[3pt]

2024 &
\textit{Programmable Polymeric-Interface for Voiceprint Biometrics}
(Voice biometrics / sensing) \citep{babu2024programmable} &
States that generating a ``unique individual voice signature'' remains a technical challenge and proposes a voiceprint-based biometric representation. \\

\addlinespace[3pt]

2024 &
\textit{Residual Networks for Text-Independent Speaker Identification: Unleashing the Power of Residual Learning}
(Speaker identification) \citep{gambhir2024residual} &
Describes speaker-identification systems as distinguishing individuals through ``unique acoustic patterns'' in their spectrograms, attributed to anatomical and behavioral differences. \\

\addlinespace[3pt]

2024 &
\textit{Text-Independent Voiceprint Recognition via Compact Embedding of Dilated Deep Convolutional Neural Networks}
(Voiceprint recognition) \citep{karthikeyan2024voiceprint} &
Describes speaker identification as identifying individuals from their ``unique speech attributes'' and trains the proposed model to extract features characterized as unique to each speaker. \\

\addlinespace[3pt]

2025 &
\textit{DKSCNN: Deep Kronecker Siamese Convolutional Neural Network Enabled Speaker Identification}
(Speaker identification) \citep{chinnasamy2025dkscnn} &
Describes deep-learning speaker-identification methods as using ``unique voice patterns'' as training data from which speaker models are learned. \\

\addlinespace[3pt]

2025 &
\textit{Partial Fake Speech Attacks in the Real World Using Deepfake Audio}
(Speaker recognition / biometric security) \citep{alali2025partial} &
Describes automatic speaker-identification systems as analysing acoustic characteristics to construct a ``unique user voiceprint or template'' for speaker identification. \\

\bottomrule
\end{tabularx}
\end{table*}

More importantly, contemporary technological research on speaker recognition, voice biometrics, and related fields continues to describe voices, voice patterns, or voiceprints as unique characteristics of individual speakers (see Table~\ref{tab:voice_uniqueness} for illustrative examples of vocal-uniqueness language in contemporary technical literature). In such work, the concern is not merely conceptual. There has been a substantial body of research showing that speaker embeddings or other voice biometrics encode not only speaker-related information, but also other nuisance and condition-specific information \citep{luu2020channel, zhao2022probing, peri2023study, baali2025sveritas}.

As a result, systems that perform very well on matched benchmarks may degrade substantially under out-of-domain conditions. Speaker representations should therefore not be interpreted as stable or unique identity codes solely on the basis of benchmark performance, and their robustness should instead be evaluated across relevant sources of variability and domain mismatch, as discussed later in this review.

The more fundamental question, therefore, is whether research built on assumptions of vocal uniqueness and stability rests on scientifically justified premises in the first place. If speaker-recognition research seeks, models, or interprets a "unique voiceprint" as an intrinsic property of an individual, a prior question must be addressed: does such a stable and individually unique vocal pattern actually exist? 

When the language of voiceprints is incorporated into technological systems, policy frameworks, and scientific discussions, it can encourage the treatment of a variable speech signal as if it were a fixed imprint of the individual. This is precisely where the voiceprint metaphor becomes misleading and potentially dangerous in practice. The voiceprint concept gains plausibility by combining well-founded observations that voices contain speaker-related information, with a much stronger and scientifically unsupported claim that each person possesses a unique, stable, and persistent vocal biometric\footnote{Although the article does not assume that every use of the term explicitly endorses all of these claims, it argues that the term invites an imprint-like interpretation that is inconsistent with the evidential nature of speech.}. It is this inference, rather than the use of speech for speaker recognition or forensic comparison, that constitutes the fallacy.

\begin{figure}[!t]
  \centering
  \includegraphics[width=\linewidth]{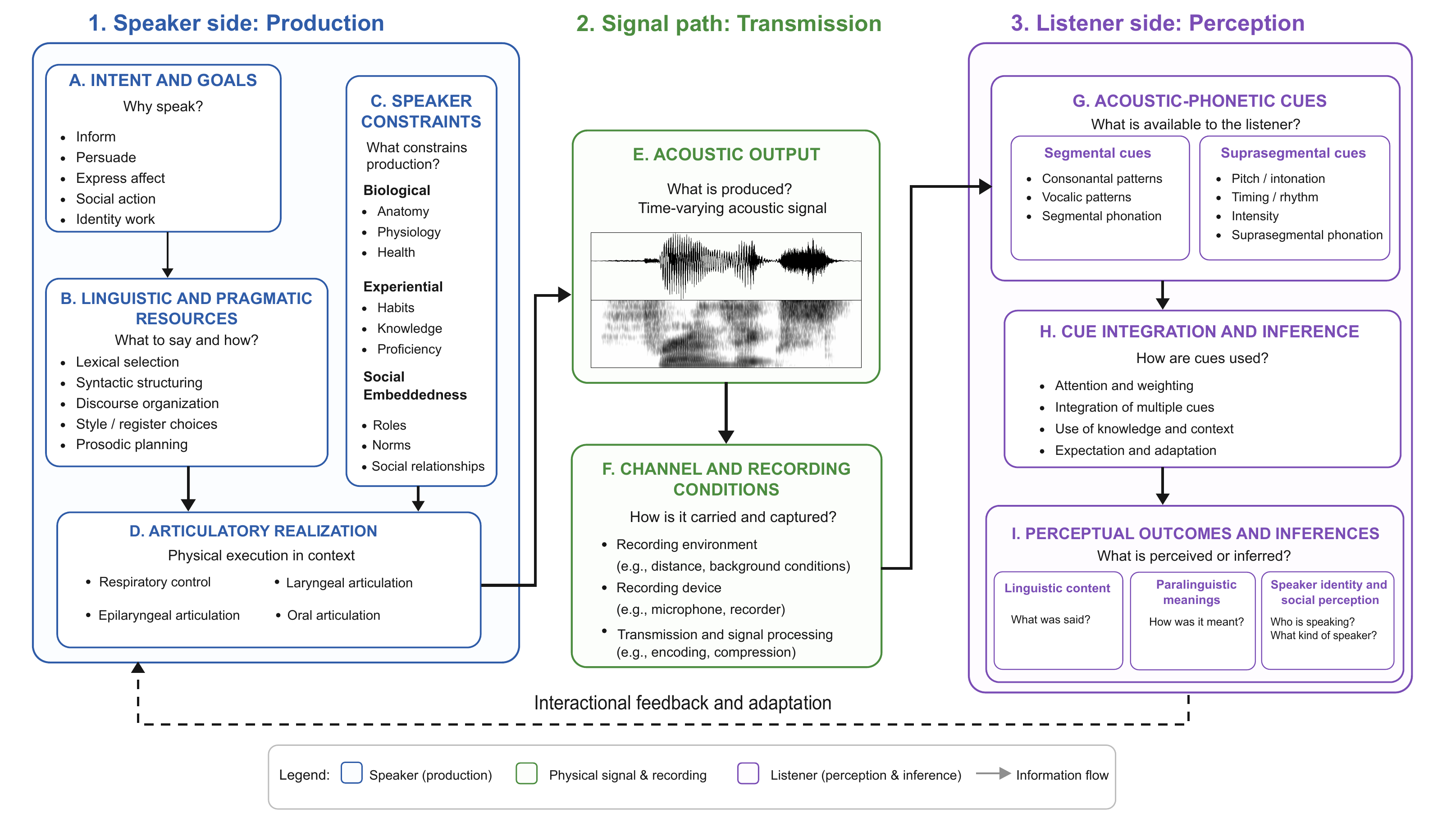}
  \caption{A model of the speech chain for recorded voice evidence.}
  \label{fig:voice_model}
\end{figure}

Voices are not static marks waiting to be recovered and matched. In Nolan's semiotic account, a voice arises from the interaction between “constraints” imposed by the physical properties of the vocal tract and “choices” made through the linguistic system in pursuit of communicative goals \citep{nolan1997speaker}. The speech signal is therefore produced anew in each act of speaking, and the speaker-related information it carries is not a fixed imprint reproduced intact across recordings, but a context-dependent pattern realized through the interaction of linguistic, physiological, social, and recording conditions. Figure \ref{fig:voice_model} summarizes this process, drawing on broader accounts of speech production, transmission, and perception \citep{nolan1983phonetic, lindblom1990explaining, scherer2003vocal}.

In this paper, we examine the scientific basis of the assumption that individuals possess unique voiceprints. We explain how voices vary under different conditions, why their measurable features vary, and what human and automatic speaker-recognition systems can and cannot establish about speaker identity. We summarize evidence from speech science, forensic phonetics, speaker recognition, and voice biometrics, while using the historical development of forensic voiceprinting to trace the origins and consequences of the uniqueness assumption. 

Our central contribution is to distinguish speaker-related information and empirical discriminability from the stronger claim of individual vocal uniqueness. High recognition performance under a given evaluation condition shows that speech can support effective speaker discrimination in that setting; it does not establish the existence of a stable, population-unique vocal pattern. This distinction has direct implications for speaker-recognition and voice-biometric systems: learned representations should be interpreted and validated with respect to relevant sources of variability, mismatch, and manipulation, including linguistic content, speaking style, physiological and affective state, age, recording and channel conditions, speech duration, and synthetic speech. Rather than treating a speaker representation as a fixed identity code, systems should evaluate their robustness to such factors and limit identity-related claims to the conditions under which performance has been empirically established.

\section{Review Scope and Search Strategy}
This article presents a narrative review of literature published between 1943 and 2026. The review begins with the invention of voiceprint (spectrographic) identification and traces the subsequent development of research on human voice variability, automatic speaker recognition, forensic voice comparison, and speech deepfakes.

Relevant publications were identified through searches of Web of Science, Scopus, PubMed, Google Scholar, the ISCA Archive, and ScienceDirect. Google Scholar was used primarily to identify older publications and to conduct backward and forward citation tracing. Searches combined terms from the following five thematic areas:

\begin{enumerate}
\item Historical voiceprint (spectrographic) identification.
\item Human voice variability and speaker individuality.
\item Forensic voice comparison and evidential interpretation.
\item Automatic speaker recognition and voice biometrics.
\item Voice cloning, voice conversion, and speech deepfakes.
\end{enumerate}

Included sources comprise historical analyses, empirical studies, forensic guidelines, validation studies, and technical reviews. Particular attention is given to studies addressing within-speaker variability, between-speaker variability, validation procedures, error rates, and evidential interpretation.

\section{The Historical Rise of Voiceprint Identification}
\subsection{The Birth of the Voiceprint}
The history of voiceprinting begins with the invention of the sound spectrograph, an instrument developed at Bell Telephone Laboratories to transform speech into a visual record of acoustic energy across time and frequency \citep{koenig1946sound}. For the first time, a spoken utterance could be preserved on paper, inspected after the sound had disappeared, and compared with other recordings. The technology was publicly introduced through the Visible Speech program, which emphasized speech analysis, pronunciation training, and the possibility of making speech visually accessible to deaf users \citep{potter1947visible}. For many years, this was the familiar account of how the sound spectrograph had developed. It was accurate, but incomplete.

A less public history remained hidden in classified wartime records. During the Second World War, Bell Laboratories researchers investigated whether spectrograms could be used to identify speakers intercepted through enemy radio communications. A 1943 draft was titled "Voiceprint Method of Identifying Individuals," and C.H.G. Gray and G.A. Kopp completed a classified report titled "Voice Print Identification" in 1944. These documents remained classified until 1960 and were largely absent from later accounts of the origins of voiceprinting. Their eventual recovery from archival collections revealed that the wartime project was not a primitive version of the method later promoted by Lawrence Kersta. It was, in several respects, considerably more cautious and methodologically developed \citep{braun2019visible}.

According to material released by \cite{braun2019visible}, Gray and Kopp did state that a voice print might identify a speaker, but the report did not present this as an established fact. It described the investigation as preliminary and repeatedly acknowledged that substantial further work was required. The researchers understood that the same speaker could produce different acoustic patterns because of transmission conditions, microphones, speaking situations, emotional states, ordinary variation across occasions, and deliberate disguise. Voice was therefore treated not as a fixed anatomical mark, but as the acoustic result of habitual speech behavior that remained open to variation.

The proposed procedure also involved considerably more than placing two spectrograms side by side and deciding whether they looked alike. Gray and Kopp described the examination of respiration, phonation, resonance, articulation, pitch, rhythm, speech rate, pauses, vowels, consonants, and other acoustic characteristics. They distinguished conspicuous vocal features from the more difficult task of separating speakers whose voices contained no obvious abnormality. In such cases, they argued that several characteristics would need to be measured and interpreted together. Their aim was not to discover a single visible mark belonging exclusively to one person, but to compare a collection of features that might jointly distinguish one speaker from another.

The report also proposed the construction of reference files containing recordings and spectrograms from known enemy speakers. New intercepts could be compared with these files, after which the closest visual match would be checked by listening to the original recordings. The researchers further suggested collecting information about how frequently particular speech characteristics occurred in relevant speaker populations. They anticipated that such data might eventually support a probability table, since a similarity shared by many speakers would be less informative than one that was relatively uncommon. Although this proposal was not developed into a validated statistical method until the rise of forensic voice comparison, it shows that the researchers at Bell Lab did not regard similarity itself as proof of identity.

The currently released archival record does not indicate that this procedure was successfully tested or used in military operations. The report provided no error rates, blind trials, or controlled evaluation of identification accuracy. A planned experiment involving repeated recordings and vocal disguise might never have been completed, and there is no evidence that the proposed trial group for testing speaker identification under field conditions was ever established. The wartime project therefore remained an ambitious program rather than a demonstrated identification technique. When the war ended, the classified work remained largely unseen from public view, while the educational account of Visible Speech became visible.

\subsection{The Rise of Voiceprint Identification}
In the early 1960s, the term "voiceprint" reappeared in a different form and came to public attention through the promotion of \cite{kersta1962voiceprint}. Kersta did not originate the idea of identifying speakers from spectrograms. His contribution was to replace the complex wartime proposal with a simpler and more easily communicated procedure. Spectrograms of the same ten frequently occurring words from questioned and known recordings were placed beside one another and judged primarily through visual comparison. Listening was no longer treated as necessary, and the earlier concern with recording conditions, speaker variation, multiple acoustic measurements, and population frequencies was largely removed. Strikingly high reported success rates further reinforced the reliability of this simplified method. Kersta reported a correct-identification rate of 99.65\% in closed-set experiments using contemporary recordings and isolated clue words \citep{kersta1962voiceprint}. This simplification also changed the role of the fingerprint analogy. In the wartime report, fingerprints and handwriting were invoked to describe visual records that could be examined and filed as reference samples. Kersta retained this analogy but placed much greater weight on the visual matching of spectrograms as the basis of identification. The spectrogram was therefore treated more as a feature used to identify the speaker.

By the early 1970s, voiceprint evidence had begun to gain judicial acceptance, with courts admitting it in criminal proceedings and, in Florida, allowing it to corroborate other identification evidence \citep{engen1972evidence,senn1973voiceprints}. This legal acceptance helped blur the distinction between speech and its visual representation. The historical shift was therefore not simply always an improvement. It was a shift from a cautious proposal for comparing variable speech characteristics to the much stronger claim that a voice imprint could function as a unique mark of speaker identity.

The next stage in the history of voiceprinting was therefore attempts to establish the reliability of such a method. A large study conducted at Michigan State University trained 29 examiners and evaluated 34,996 visual comparisons under open and closed conditions, using recordings made at the same or different times and several forms of verbal material. In conditions designed to approximate forensic comparisons, forced decisions resulted in false-identification and false-elimination rates of approximately 6\% and 13\%, respectively. The authors estimated that, if examiners had instead been permitted to report no opinion when uncertain, these error rates would have fallen to approximately 2\% and 5\%, while only 74\% of the comparisons would have produced a decision \citep{tosi1972experiment}. A later survey of 2,000 comparisons conducted by FBI examiners reported false-identification and false-elimination rates of 0.31\% and 0.53\%, respectively, but decisions were reached in only 34.8\% of the comparisons \citep{koenig1986spectrographic}. These findings became one of the principal empirical foundations for claims that spectrographic identification could be used in forensic practice.

\subsection{The Fall of the Voiceprint}
However, these studies did not resolve the underlying scientific dispute. A committee of speech scientists argued that the fingerprint analogy remained inappropriate because spectrograms did not behave like fixed individual patterns \citep{bolt1970speaker}. The same speaker could produce visibly different realizations of the same utterance, while different speakers could produce similar patterns. This committee also emphasized that error rates varied substantially with the task, the speech material, and the experimental conditions \citep{bolt1970speaker}. This dispute soon became part of the legal controversy surrounding the method. A federal appellate court later warned that the term itself gave spectrographic identification an appearance of certainty and accuracy that the available evidence did not justify \citep{baller1975}.

These disagreements also produced conflicting legal decisions. In People v. King, Kersta presented spectrographic voice identification as possessing an accuracy comparable to fingerprint identification. The appellate opinion records that his claims were opposed by seven defense witnesses drawn from relevant scientific and technical fields. Louis J. Gerstman, a former colleague who had monitored Kersta's work at Bell Telephone Laboratories, testified that identifying speakers from spectrograms was extremely unlikely to be possible, that Kersta's tests were neither reliable nor valid in the technical scientific sense, and that the method could produce a large number of errors or misidentifications. Frank Clarke testified that Kersta's method was unreliable and that he knew of no one who could verify Kersta's results. Martin Joos stated that the available technology did not permit positive identification from a spectrogram and that he knew of no authority other than Kersta who accepted the validity of the method. Peter Ladefoged characterized Kersta's tests involving voice mimics as resembling parlor tricks and warned that the courts were being asked to accept something akin to palmistry. Victoria Fromkin testified that Kersta's tests were not scientifically reliable and were not accepted by competent authorities in the field. Ralph Vanderslice presented experimental results showing that the same speaker could produce substantially different spectrograms under different conditions without attempting to disguise the voice. The appellate court also recorded that Kersta had admitted that his process was an art and that his opinion was entirely subjective. It consequently ruled that spectrographic voice identification had not reached a sufficient level of scientific certainty to be accepted as identification evidence in cases in which a defendant's life or liberty might be at stake \citep[pp. 442, 451 to 456, and 460]{people1968king}.

The D.C. Circuit reached a similar conclusion in United States v. Addison, finding that spectrographic speaker identification had not achieved the general scientific acceptance required under Frye \citep{addison1974}. However, other courts were more receptive. In State v. Williams, the Supreme Judicial Court of Maine admitted the evidence under a relevance and helpfulness standard, treating concerns about examiner judgment and the difference between experimental and forensic conditions as questions about evidential weight rather than admissibility \citep{state1978williams}.

The continuing disagreement led the National Research Council to review the technology, the experimental literature, and its use in court. The Council found that contemporary aural and visual identification depended heavily on limited scientific knowledge and examiner experience. It also concluded that the assumptions underlying the method, particularly those concerning variation within the same speaker and overlap between different speakers, had not been adequately supported by scientific theory or representative data. Its report therefore urged great caution in forensic applications \citep{national1979theory}. Despite these scientific concerns, spectrographic identification remained in operational use for some time, including within the FBI \citep{koenig1986spectrographic}. By the early 1990s, however, its use was continuing to decline despite attempts to introduce new procedural standards \citep{koenig1993selected}.

By the late 1990s, voiceprinting as a distinct forensic method was drawing to a close, but forensic speaker recognition was entering a new phase. The field increasingly adopted various auditory, acoustic, and computer-assisted approaches that treated speaker comparison as a broader evidential problem rather than as an improved form of visual spectrogram matching \citep{kunzel1994current}. This transition also shifted the scientific debate away from the apparent resemblance between spectrograms and fingerprints and toward empirical validation, reliability under relevant conditions, and the defensible interpretation of evidential strength \citep{morrison2014distinguishing}. In this sense, the decline of voiceprinting marked the emergence of contemporary forensic voice comparison, which is discussed in Section \ref{sec:FVC}.

\section{Human voice variability and speaker individuality}
\label{sec:variability}
This section discusses human voice variability and speaker individuality. The first subsection examines within-speaker variability and explains why a speaker’s voice should not be treated as a fixed acoustic object. The second subsection considers between-speaker variability and the problem of overlap, showing why speaker differences support probabilistic comparison rather than imprint-like individualization.

\subsection{Within-Speaker Variability}
\subsubsection{Short-Term and Situational Variability}
Within-speaker variability is partly driven by short-term changes such as the speaker's affective, physiological, and communicative state. These changes can affect multiple levels of the speech signal, including phonation, articulation, timing, intensity, prosody, and voice quality. They therefore need to be treated as ordinary sources of voice variation rather than as marginal exceptions.

Emotional state is one well-established source of such variability. A classic acoustic study of vocal emotion expression showed that different portrayed emotions were associated with systematic differences in vocal parameters, including pitch, intensity, timing, and voice quality \citep{banse1996acoustic}. The review of \cite{scherer2003vocal} further framed emotional speech as a process in which affective states are encoded through multiple acoustic cues rather than through a single invariant marker. Evidence in Mandarin further shows that emotional vocalization is associated with both spectrum-based and F0-related cues \citep{wang2024acoustic}, and it was similarly shown that emotional effects are distributed across multiple acoustic dimensions rather than localized in a single cue \citep{wang2024gender}. Recent review work again shows that affective state can be reflected in prosodic, spectral, and voice-quality measures, while also emphasizing substantial heterogeneity across studies and tasks \citep{schewski2025measuring}. These studies show that emotional variation can reshape the acoustic realization of the same speaker across recordings.

Stress and cognitive load provide a related but distinct source of variation. Work on speech under stress has shown that stress-related conditions can affect acoustic properties such as fundamental frequency, duration, intensity, and spectral characteristics \citep{hansen1996analysis}. For example, a study modeled drivers' speech under varying levels of cognitive load and showed that stress-related speech can be classified using acoustic features derived from the speech signal \citep{fernandez2003modeling}. Similarly, in an aviation simulator study, it was found that increased cognitive load was associated with higher mean fundamental frequency and vocal intensity, as well as a reduced fundamental frequency range \citep{huttunen2011effect}. Work on cognitive load and autonomic arousal during voice production found that increased cognitive load was associated with increased sound pressure level, while autonomic arousal was related to changes in cepstral peak prominence \citep{tomassi2025investigating}. These findings indicate that temporary task demands and psychological pressure can alter the acoustic form of a speaker's voice.

Fatigue introduces another source of short-term within-speaker variability. Vocal fatigue has been described as a multidimensional phenomenon involving increased perceived phonatory effort, reduced vocal function, and possible changes in laryngeal control \citep{welham2003vocal, solomon2008vocal}. More recent work distinguishes vocal fatigue from related concepts such as vocal effort, vocal load, and vocal loading, emphasizing that these terms refer to connected but not identical aspects of vocal demand and vocal response \citep{hunter2020toward}. Fatigue can also arise from broader physiological states. For instance, sustained wakefulness has been shown to affect speech timing and frequency-related acoustic measures \citep{vogel2010acoustic}. These studies show that a speaker's voice may vary not only because of how they choose to speak, but also because of changing physical capacity and vocal-system demand.

Health state and hydration can also affect voice production. Work on vocal fold physiology shows that hydration influences the biomechanical properties of the vocal folds and that dehydration can increase the effort required for phonation \citep{sivasankar2010role}. Experimental work on hydration and voice quality further suggests that systemic dehydration and surface dehydration can affect measures such as perturbation, noise-related measures, and maximum phonation time \citep{alves2019effect}. Respiratory and viral illness can introduce additional changes because the articulatory and respiratory systems are directly involved in speech production. Studies of respiratory infection and COVID-19 have reported changes in voice quality and acoustic speech characteristics associated with illness \citep{bartl2021voice, talkar2023dissociating, berti2025acoustic}. Mental health is another important source of variation. For example, depression has been associated with differences in speech timing, prosody, and voice quality, although the specific patterns reported vary across studies and speaking contexts \citep{almaghrabi2023bio, hall2024phonetic}. Health-related variation therefore affects the speech signal through both vocal mechanisms and broader respiratory or articulatory constraints.

Communicative demands, audience orientation, and self-monitoring also shape within-speaker variability. Sociolinguistic work on attention to speech and audience design shows that speakers shift styles in relation to monitoring, listeners, settings, and interactional goals \citep{labov1973sociolinguistic, bell1984language, sharma2018style}. Experimental work on clear speech further shows that speakers modify acoustic properties such as speaking rate, segmental articulation, and intensity when asked to speak more clearly \citep{picheny1986speaking, lam2012acoustics}. Infant-directed speech provides a further example of audience-specific modulation: when speaking to infants, caregivers commonly raise their mean fundamental frequency and use more variable or exaggerated pitch contours, often alongside changes in temporal organization such as shorter utterances, longer pauses, and reduced global speech rate \citep{fernald1985four, fernald1989cross, soderstrom2007beyond, narayan2016speech}. Speech produced in noise provides another example of situationally conditioned variation, with Lombard speech involving changes in vocal intensity, fundamental frequency, spectral structure, and timing \citep{junqua1996influence, uma2021understanding}. Together, these lines of work show that short-term changes in task, audience, environment, and self-monitoring can produce systematic acoustic differences within the same speaker.

\subsubsection{Linguistic and Stylistic Conditioning}
Speaker-related acoustic information is always realized through particular linguistic material and speaking conditions, and therefore cannot be assumed to be observed independently of what is said and how it is said. Segmental composition shapes the acoustic signal because vowels and consonants differ in formant structure, duration, spectral properties, and articulatory configuration \citep{peterson1952control, hillenbrand1995acoustic}. Even within the same vowel category, acoustic realization may shift with duration and surrounding segmental context \citep{lindblom1963spectrographic, zhang2020effect, havenhill2024articulatory}.

Speaking style further conditions how speaker-related information appears in the signal. Multilingual work on scripted and spontaneous speech classification shows that speaking style is reflected in acoustic and prosodic patterns across languages, while also indicating that these patterns may vary across datasets and language groups \citep{elisha2024classification}. Work on degrees of spontaneity further suggests that read and spontaneous speech should not be treated as a simple binary contrast, because style categories can overlap and contain internal variation \citep{adigwe2025modelling}. Speaker-dependent vowel characteristics also vary across speech styles, suggesting that speaker-indexical information is conditioned by the material and style through which it is produced \citep{audibert2024speaker}. Listener-oriented speaking style and semantic predictability also affect vowel production, including vowel duration, dispersion, and F2 \citep{song2024effects}.

\begin{comment}
The acoustic differences among speech styles are also reflected in their lexical distributions. In work on homophone duration, spontaneous corpora were found to contain more high-frequency lexical items, which tend to undergo frequency-related reduction, and this lexical-compositional difference accounted for a substantial part of the duration difference between read and spontaneous speech \citep{yang2026homophone} \textcolor{red}{Cite later}.
\end{comment}

Prosodic structure adds another layer of conditioning. Fundamental frequency, duration, and intensity are also shaped by intonational structure, prominence, boundary marking, and discourse function \citep{pierrehumbert1980phonology, ladd2008intonational}. Prosodic prominence is realized through combinations of acoustic cues, and the contribution of these cues can differ by vowel category. For example, duration has been found to be the strongest overall predictor of prominence ratings, while the relative roles of fundamental frequency, intensity, formants, and tones vary across vowels \citep{zhang2023contribution}. Focus and constituency also shape prosodic realization through prominence and phrasing, and these functions may be encoded simultaneously with language-specific cue weighting \citep{zhang2024prosodic}. Phrase position also conditions prosodic realization, with Singapore English showing fundamental frequency rises toward phrase-final syllables and additional duration and intensity evidence for phrase-final prominence \citep{chong2023prominence}. Therefore, acoustic differences across samples may arise from differences in linguistic organization and discourse context, not only from differences between speakers. Voice samples are not context-free measurements of a speaker's voice, but samples of a particular speaker producing a particular linguistic material in a particular style.

\subsubsection{Long-Term and Life-Course Variability}
Some sources of within-speaker variability unfold over much longer time scales. A speaker's voice changes across the life course as the vocal folds, vocal tract, respiratory system, linguistic repertoire, and patterns of voice use change over time \citep{reubold2010vocal, rojas2020does, haddad2024presbyphonia}. A famous longitudinal example comes from Queen Elizabeth II's annual Christmas broadcasts, which showed substantial changes in vowel quality between the 1950s and the late 1960s and early 1970s, followed by comparatively little change through the 1980s \citep{harrington2000monophthongal}. These changes shifted her vowels toward a more mainstream form of pronunciation, demonstrating that an adult speaker's phonetic patterns can change across time. For this reason, unlike fingerprints, which are often treated as relatively persistent anatomical patterns \citep{maltoni2009handbook, yoon2015longitudinal}, a voice sample recorded at one point in a speaker’s life cannot be treated as a permanent acoustic record of that speaker.

Aging is one major source of long-term voice change. Age-related changes in the larynx, vocal fold tissue, respiratory support, and vocal tract can affect fundamental frequency, voice quality, formant structure, and articulatory control \citep{sataloff2005effects, lortie2015effects, mallick2019presbylaryngis}. Longitudinal studies further show that speakers may exhibit measurable changes in fundamental frequency and formant frequencies over periods of decades, even when the comparison is made within the same individuals rather than across age groups \citep{harrington2007age, reubold2010vocal}. Cross-sectional and clinical work similarly indicates that healthy aging and age-related dysphonia do not produce a single uniform acoustic profile, but can affect multiple voice dimensions in different ways \citep{schultz2023cross, cavallaro2024exploring}.

Hormonal and developmental changes also shape the voice across the life course. Pubertal development is associated with large changes in vocal fundamental frequency and vocal tract dimensions, while later hormonal changes may affect vocal fold tissue, pitch, vocal range, and perceived voice quality \citep{harries1998changes, zamponi2021effect, fujiki2025pediatric}. Work on menopause and hormone-related voice change further suggests that changes in endocrine state can be associated with lower fundamental frequency, vocal instability, fatigue, or reduced vocal range, although the size and direction of these effects may vary across speakers \citep{afsah2024effects, shah2025effects}. These findings show that voice change is not limited to old age, but is tied to biological development and physiological change across adulthood.

Long-term health, vocal damage, medical treatment, and patterns of voice use provide another source of change. Voice disorders such as presbyphonia involve changes in vocal fold closure, pliability, and vocal function, and treatment studies show that voice quality and function can be modified through therapy and behavioral intervention \citep{saccente2024systematic, haddad2024treating}. Repeated vocal loading and phonotrauma can affect vocal fold tissue and contribute to lesions, while systemic and psychosocial conditions can alter respiration, laryngeal tension, fatigue, and voice quality \citep{dietrich2008frequency, galindo2017modeling, misono2014psychosocial}.

The linguistic and social aspects can also reshape speech over time. Bilingual and second-language research shows that new language experience can affect first-language speech production, including voice onset time, vowel realization, and cross-language phonetic category structure \citep{dmitrieva2020effect, osborne2021foreign, turner2023phonetic}. Longitudinal work on second-dialect acquisition similarly shows that relocation and dialect contact can shift some vocalic patterns over many years \citep{cheng2023second}. These studies show that linguistic experience can reshape first-language speech production over time. A particularly direct example of deliberate vocal change is provided by gender-affirming voice training. Studies of transgender women show that structured training can alter fundamental frequency and formant patterns, as well as how the voice is perceived by speakers themselves and by listeners, helping speakers develop vocal patterns that are more congruent with their gender identity and communicative goals \citep{leyns2023short, oates2023gender}. These findings suggest that the voice is partly learned and can be continually updated through linguistic experience and deliberate practice.

Long-term variability is therefore central to the problem of speaker identity. Some speaker-specific vocal tendencies may persist across time, but persistence does not imply acoustic invariance. A voice sample is a time-bound realization of those tendencies under particular biological, linguistic, and social conditions, not a fixed biometric record that can stand for the speaker across the life course.

\subsubsection{Intentional Manipulation}
Unlike fingerprints, voice evidence can be intentionally manipulated by the speaker through changes in pitch, phonation, articulation, resonance, accent, or speaking style in order to conceal, modify, or project a different vocal identity \citep{masthoff1996report, perrot2007voice, eriksson2010disguised}. Changes in speaking fundamental frequency are a common strategy, and speakers may raise or lower their pitch, shift into falsetto, produce creaky voice, or otherwise modify their phonatory setting \citep{kunzel2000effects, kunzel2004effect}. Such pitch manipulation may produce accompanying changes in syllable duration, intensity, vowel formants, and long-term spectral characteristics, and may consequently affect human and automatic speaker recognition \citep{zhang2008voice, zhang2012acoustic}.

Other forms of disguise affect voice quality, resonance, and vocal tract configuration, including whispery voice, nasal modification, and age-related voice imitation \citep{san2013civil, zhang2017acoustic, hautamaki2017acoustical}. In Mandarin, whispery disguise has been shown to affect articulation rate, intensity, and vowel formant frequencies, while also creating difficulties for human and automatic speaker recognition \citep{zhang2018whispery}. Accent alteration is another strategy through which speakers may reshape speaker-indexical information. Attempts to perform a foreign or socially marked accent can affect segmental realization, timing, and prosody because accent is distributed across multiple phonetic dimensions \citep{silva2023voice}.

Impersonation is a related but distinct form of intentional manipulation: rather than simply concealing identity, it attempts to approximate another speaker. Acoustic and perceptual work suggests that non-expert speakers can deliberately alter their vocal output in impersonation tasks, even if the magnitude of these acoustic changes is limited \citep{delvaux2017voice}. Because impersonation may involve changes in speech style, prosody, accent, and socially meaningful speaker traits, it remains a difficult case for treating voice samples as stable speaker evidence \citep{gu2024utilizing}.

However, intentional manipulation does not imply that speakers can change every aspect of their voice equally well. Some dimensions may be easier to control, while others may remain partly constrained by anatomy, long-term motor habits, language background, or limits on vocal flexibility \citep{delvaux2017voice, hautamaki2017acoustical, gu2024utilizing}. The important point here is that voice samples can be affected by deliberate speaker manipulation. A speech recording therefore cannot be assumed to preserve a stable and unmodified biometric object, because a speaker may actively participate in changing the acoustic evidence.

\subsection{Between-Speaker Variability and the Problem of Overlap}
The existence of within-speaker variability does not mean that speakers are indistinguishable. Voices contain speaker-related information because speakers differ in anatomical, physiological, linguistic, and habitual properties, and such differences provide the basis for both human speaker recognition and automatic speaker recognition \citep{nolan1983phonetic, kinnunen2010overview, afshan2020speaker}. The problem here, however, is not whether speakers differ, but how those differences are distributed. A speaker should not be understood as a single fixed acoustic point. The same speaker may produce a range of acoustic patterns across contexts, while different speakers may occupy partially overlapping regions of acoustic space. A useful speaker-related feature is therefore not simply one that differs across speakers, but one whose between-speaker variability is large relative to its within-speaker variability under relevant conditions \citep{leemann2014speaker}.

This distributional view makes overlap central to speaker individuality. Empirical work on forensic voice comparison shows that the strength of speaker discrimination depends on what linguistic material is sampled, since phonemic content can affect both intra-speaker and inter-speaker variability \citep{moez2016phonetic}. Comparisons across speaking styles also show that speaker-discriminatory power is not symmetric across materials: speakers may be more or less separable depending on whether spontaneous dialogues or interviews are used \citep{cavalcanti2023speaker}.

Technical mismatches create a related problem because they can reduce the observable separation between speakers: differences in recording channel, microphone condition, telephone transmission, or speaker-to-device distance may degrade the speaker-discriminative information available in the signal and increase overlap between same-speaker and different-speaker comparisons \citep{alexander2004effect, zhang2013effects, carne2015likelihood, enzinger2015mismatched, van2020exploring}. Automatic speaker-verification systems show the same issue at the score level: acoustic mismatch between enrollment and test speech can reduce target-speaker scores, making speaker separability dependent on the conditions under which the samples are compared \citep{hautamaki2020did}. Similarity therefore cannot by itself establish identity, and difference cannot by itself exclude identity.

This is why speaker differences support probabilistic comparison. Voice evidence is meaningful because speakers are not all the same, but its application depends on the degree of overlap between same-speaker and different-speaker distributions. In this sense, the relevant question is not whether two recordings “match,” but how much more probable the observed evidence is under one speaker-related proposition than under an alternative proposition \citep{morrison2009forensic, hughes2017relevant, morrison2021consensus}. Speaker individuality is therefore a statistical relationship between within-speaker variation and between-speaker variation, not proof of a unique and stable vocal imprint.

\section{Forensic Voice Comparison}
\label{sec:FVC}
\subsection{Current Practice Across Jurisdictions}
Contemporary forensic voice comparison is not a single, globally standardized practice. A survey of law enforcement agencies found substantial differences across countries and institutions, with responding agencies using spectrographic comparison, auditory analysis, auditory-acoustic-phonetic analysis, semi-automatic systems, automatic systems, or combinations of these methods \citep{morrison2016interpol}. The rejection of voiceprinting has therefore not produced an immediate or uniform transition to a common alternative.

China provides a particularly clear example of this uneven development. The Ministry of Public Security standard GA/T 1433-2017 describes voice identification as the examination, comparison, and comprehensive assessment of "voiceprint features." Its procedure combines auditory analysis with qualitative and quantitative spectrographic comparison, the requirement of certain syllables, formants, agreement and difference, followed by one of five conclusions: identification as the same speaker, towards identification as the same speaker, exclusion of the same speaker, towards exclusion of the same speaker, and inconclusive \citep{ministrypublicsecurity2017}. The Ministry of Justice standard SF/T 0122-2021 retains a broadly similar structure based on auditory features, spectrographic features, and categorical or qualified conclusions \citep{ministryjustice2021}, but without the concrete quantitative requirement of syllables, formants, agreement and difference. Both standards avoid using the term "voiceprint identification", but the term is still widely used in practice. At the same time, Chinese forensic scholarship has explicitly criticized traditional voiceprinting and developed likelihood-ratio approaches based on quantitative measurements, relevant population data, statistical modelling, and validation under conditions reflecting the case \citep{zhang2016sisters, zhang2023paradigm}. Chinese practice therefore contains both older individualization language and newer models of probabilistic evidence evaluation.

In the United States, the situation is similarly fragmented, although the institutional structure differs from that in China. At the federal level, expert evidence is assessed under Federal Rule of Evidence 702, which requires the proponent to establish that the testimony is based on sufficient facts or data, reliable principles and methods, and a reliable application of those methods to the case, but does not prescribe a method or reporting framework specific to forensic voice comparison \citep{federalrule702}. The recent OSAC guide also states that no single method is used by all practitioners and does not promote any one method of analysis or interpretation framework over another \citep{osac2024}. The guide therefore describes the range of approaches used within the field without establishing a uniform procedure.

Within Europe, ENFSI has developed a more explicit shared professional framework. Its guideline for evaluative reporting sets out a framework in which findings are evaluated under competing propositions and evidential strength is expressed through a likelihood ratio \citep{willis2015guideline}. An ENFSI-supported methodological guideline for semiautomatic and automatic speaker recognition places case assessment, validation, interpretation, reporting, and proficiency testing within a Bayesian likelihood-ratio framework \citep{drygajlo2015methodological}. For forensic speaker comparison based on combined phonetic-linguistic auditory and acoustic analysis, the ENFSI Best Practice Manual provides guidance on practitioner competence, examination, validation, peer review, quality assurance, and reporting. It also acknowledges that the choice of methodology depends on the material, the techniques available within the laboratory, and the rules of the relevant jurisdiction, and that no conclusion scale is universally used \citep{enfsi2022speaker}. ENFSI therefore provides its member laboratories with a coordinated professional framework, but not a single method uniformly used throughout Europe.

Australia and New Zealand represent another regional approach. The National Institute of Forensic Science within the Australia New Zealand Policing Advisory Agency (ANZPAA NIFS) has issued cross-disciplinary guidance on evaluative reporting, under which findings are assessed under competing propositions and evidential strength may be expressed as a numerical likelihood ratio or a verbal equivalent \citep{anzpaa2017evaluative}. Likelihood-ratio-based forensic voice comparison has also been used in an Australian trial \citep{rose2013where}, and ANZPAA NIFS has issued general guidance on method validation \citep{anzpaa2025validation}. These materials provide shared cross-disciplinary guidance on evaluative reporting and method validation across Australia and New Zealand, but do not prescribe a uniform method specifically for forensic voice comparison.

More traditional approaches to speaker identification remain visible in some contemporary official practice and have also been reported across several regions. In India, the Uttar Pradesh Police describes forensic audio analysis as using a “Spectrographic Analysis Method for Speaker Identification,” while the Maharashtra Directorate of Forensic Science Laboratories lists spectrography among the techniques used for speaker identification and characterizes the voice as carrying unique characteristics “akin to a fingerprint” \citep{uppolice_speakeridentification,maharashtra_speakeridentification}. A similarly explicit approach is described by Japan’s National Research Institute of Police Science, which presents “individual identification by voiceprint” as the comparison of voiceprints obtained through computerized frequency analysis of offender and suspect recordings\footnote{Descriptions of current institutional practices reflect information available on the cited official websites at the time of access; web content may subsequently be revised, relocated, or removed.} \citep{japan_nrips_voiceprint}.

At a broader regional level, an INTERPOL survey that received 91 responses from 69 countries reported that spectrographic or auditory-spectrographic approaches were the most commonly reported approaches among respondents in Africa, Asia, the Middle East, and South and Central America \citep{morrison2016interpol}. The same survey nevertheless emphasized substantial diversity in speaker-identification methods and reporting frameworks across regions \citep{morrison2016interpol}. A subsequent survey of 39 forensic speaker-comparison respondents across 23 countries similarly documented considerable variation in practice: 41.2\% reported using automatic speaker recognition, all respondents using such systems also reported combining them with human analysis, and several different frameworks for expressing conclusions remained in use \citep{gold2019international}. These findings suggest that spectrographic speaker identification has not been confined to a certain country. At the same time, the available international evidence points to heterogeneous practices rather than a uniform approach across jurisdictions.

\subsection{How Forensic Voice Evidence Is Evaluated}
Contemporary forensic voice comparison should begin by defining competing source propositions\footnote{The following account is normative rather than descriptive: although current jurisdictions permit multiple approaches, the evaluative framework advocated here is proposition-based, probabilistic, and empirically validated.}. One proposition normally states that the questioned and known recordings were produced by the same speaker. The alternative states that the questioned recording was produced by another speaker drawn from a relevant population. Before any comparison is interpreted, the examiner must determine whether the recordings contain sufficient and comparable speech. Relevant considerations include recording quality, language and dialect, speaking style, channel, temporal separation, health, emotional state, and possible disguise \citep{morrison2019introduction}.

The recordings may then be examined using methods such as auditory-phonetic analysis, measurements of acoustic features, automatic speaker-comparison systems, or a combination of these approaches. The value or result depends on the degree of both similarity and typicality. A feature shared by two recordings provides little support for a common speaker if it is also frequent among other speakers in the relevant population. The same degree of similarity may provide greater support when the observed feature is uncommon. The likelihood ratio expresses the ratio of the probability of the observed evidence under the same-speaker proposition to that under the different-speaker proposition. It does not directly express the probability that the known speaker produced the questioned recording \citep{gold2014issues}.

Relevant population data, quantitative measurements, and statistical models may be used to convert acoustic features or automatic comparison scores into numerical likelihood ratios. Where the available data do not support a defensible numerical calculation, conclusions may be expressed using a verbal likelihood-ratio scale, provided that the reasoning, empirical basis, and uncertainty are made explicit. Whatever method is used, it should be validated under conditions that approximate those of the case \citep{sergidou2024fusing}. Validation should examine both discrimination and calibration and should reflect relevant differences in channel, language, duration, speaking style, and recording quality \citep{morrison2009forensic,morrison2021consensus}.

The scientifically defensible alternative to voiceprinting is therefore not a more detailed visual match or a more powerful speaker representation. It is a transparent, proposition-based, and empirically validated evaluation of how strongly the recordings support one source explanation over another.

\section{Automatic Speaker Recognition and Voice Biometrics}
Automatic speaker recognition refers to the use of computational systems to distinguish or compare speakers from their speech recordings. Speaker identification systems select the best-matching speaker from a set of candidates, whereas speaker verification systems assess whether a speech recording supports a claimed identity. In a typical voice biometric application, speech from an enrolled user is compared with a new test recording. The system then produces a numerical score that reflects the degree of similarity under its particular representation and scoring procedure \citep{kinnunen2010overview}. In practice, the system transforms a recording through several computational stages. Acoustic measurements are first extracted from the speech signal. These measurements are then represented by a statistical model or a fixed-dimensional embedding and compared with representations derived from other recordings. The output is therefore not a direct measurement of an independently existing vocal object. It is the result of a particular model operating on speech produced and recorded under particular conditions. The existence of these systems demonstrates that speech does contain measurable speaker-related information. They do not, however, establish that the voice contains a fixed identifying mark.

\subsection{From Acoustic Observations to Speaker Representations}
Earlier automatic speaker verification systems commonly represented speech using distributions of acoustic features. A good example is the Gaussian mixture model universal background model (GMM-UBM). The universal background model describes the distribution of acoustic observations in a broader population, while a target-speaker model is created by adapting this background model using the available enrollment speech. A test recording is evaluated by comparing how well it is explained by the target model relative to the background model \citep{reynolds2000speaker}. The target-speaker model is therefore estimated from limited speech data and defined relative to a large population model.

Joint factor analysis (JFA) made the treatment of speaker and channel variability more explicit by modelling them in separate low-dimensional subspaces \citep{kenny2007joint}. However, the factors intended to represent channel variation were found to retain speaker-discriminative information. This finding motivated the total-variability framework, in which speaker and channel variation are represented jointly rather than assumed to have been cleanly separated \citep{dehak2011front}. The transition is important here because even an explicit attempt to isolate a stable speaker component did not produce a context-free representation of identity.

The resulting i-vector framework introduced a more compact representation. It represents an utterance as a point in a low-dimensional total variability space. This space is learned from training data and is intended to capture the major sources of variation observed across speech recordings. Importantly, the original formulation describes total variability as including both speaker and channel variability \citep{dehak2011front}. An i-vector is therefore an estimate of where a particular recording lies within a model-derived space of variation. It does not isolate a context-free identity component before subsequent normalization, compensation, and scoring procedures are applied.

Neural speaker embeddings changed how these representations are learned. X-vector systems train a neural network to discriminate among speakers represented in the training data. Frame-level acoustic features are processed by a time-delay neural network, aggregated across the recording, and projected into a fixed-dimensional embedding. The embedding is usually taken from an internal network layer and passed to a separate scoring model \citep{snyder2018x}. Its coordinates are determined by the network architecture, training speakers, acoustic features, augmentation procedures, and optimization objective. Architectures such as Emphasized Channel Attention, Propagation and Aggregation in Time Delay Neural Networks (ECAPA-TDNN) have further improved speaker verification by expanding temporal context, modelling relationships between feature channels, aggregating information across network layers, and assigning different weights to different portions of the recording \citep{desplanques2020ecapa}.

Other neural systems use residual convolutional networks to extract ResNet-based speaker embeddings, commonly described as r-vectors \citep{zeinali2019but}. More recent systems also use self-supervised front-ends such as wav2vec 2.0 and WavLM, whose representations are pretrained on particular collections of speech before being adapted for speaker verification \citep{fan2020exploring, chen2022wavlm}. Performance can be further improved through large-margin fine-tuning, while embeddings may be compared using cosine scoring or a learned probabilistic back-end \citep{thienpondt2021idlab, wang2022scoring}. These developments improve speaker discrimination, but they do not make the representation independent of its construction. The resulting embedding remains dependent on the architecture, pretraining corpus, adaptation data, optimization objective, and scoring procedure.

The progression therefore reflects changing statistical approaches to the same basic problem. Each method attempts to emphasize variation that is useful for distinguishing speakers while reducing variation associated with other factors. None of these methods discovers a single acoustic pattern that exists independently of the recording and the model. The identity-related information used by the system is distributed across acoustic observations and is represented differently by different modelling approaches.

\subsection{Condition Dependence in Speaker Embeddings}
Speaker-discriminative training does not guarantee that an embedding contains only speaker identity \citep{heo2024rethinking}. Speech simultaneously conveys linguistic content, paralinguistic meanings, speaking style, language, and information about the recording environment. Probing experiments have shown that x-vectors retain information about spoken sentences, words, phones, channel conditions, utterance duration, and data augmentation, even while performing effectively in speaker verification \citep{raj2019probing}. A representation can therefore be useful for distinguishing speakers without providing a separate or context-free encoding of identity.

This mixed representation becomes particularly important when enrollment and test recordings come from different domains. Changes in microphone, transmission channel, language, noise environment, dataset, or recording procedure could shift the distributions of embeddings and scores \citep{yang2026domain}. The 2018 NIST Speaker Recognition Evaluation found substantial effects of domain, channel, and duration mismatch on system performance \citep{sadjadi20192018}. Subsequent evaluations examined speaker-verification systems under a broader range of conditions. SRE21 included both same-source and cross-source trials involving conversational telephone speech and audio from video, together with cross-lingual trials in which the enrollment and test segments were in different languages \citep{sadjadi2021nist}. VoxSRC evaluated speaker verification using unconstrained speech collected from YouTube videos, including professionally recorded interviews and more casual recordings containing background noise, laughter, and other artefacts \citep{chung2019voxsrc}. Analyses of such mismatched conditions have further shown that performance depends on both the embedding extractor and the scoring back-end. Language-dependent modelling, nuisance-attribute projection, and changes to the speaker subspace can substantially alter the resulting performance \citep{silnova2022analyzing}. These findings show that a representation cannot be interpreted apart from the domain in which it was trained and evaluated.

Under the VoxCeleb1-O benchmark, an ECAPA-TDNN system achieved an equal error rate of 0.87\%, while a later reproducible WavLM–ECAPA system reported an equal error rate of 0.39\% \citep{desplanques2020ecapa, jung2024espnet}. These figures demonstrate the strong performance attainable under particular benchmark conditions. As the early voiceprinting literature also illustrates, however, very low error rates obtained under controlled evaluation conditions should not be assumed to generalize unchanged to applications involving materially different data and operational conditions. These figures should therefore not be interpreted as case-specific forensic error rates, because they depend on the benchmark data, trial construction, training resources, and evaluation protocol. Their applicability to forensic casework must instead be established through empirical validation under conditions that adequately reflect the questioned- and known-speaker recordings, the speaking styles, and the relevant population in the case \citep{morrison2016multi, morrison2019multi}.

The amount and phonetic composition of the available speech affect the resulting representation, especially when enrollment or test recordings contain only a few seconds of speech. Fixed-dimensional embeddings may appear to reduce recordings of different lengths to a common identity code, but the information available to the model still depends on which sounds were produced and how often they were observed. This problem is especially apparent in short recordings. The SDSV Challenge 2020 documented the difficulty of speaker verification when enrollment and test utterances contained only a few seconds of speech \citep{zeinali2020sdsv}. An embedding extracted under such conditions is necessarily estimated from a restricted and potentially unrepresentative sample of the speaker’s behavior.

Speaking style and modality produce another source of mismatch. In an x-vector and PLDA system, verification errors were substantially lower when enrollment and test recordings shared the same style than when read, conversational, narrative, and pet-directed speech were mismatched \citep{afshan2020speaker}. Differences in vocal effort create similar problems. Shouted speech substantially degrades the performance of an x-vector speaker verification system when it is compared with normally produced enrollment speech \citep{prieto2020shouted}, while mismatch between whispered and neutral speech produces substantial performance degradation in an i-vector system \citep{naini2019whisper}.

Speaker representations can also change over longer periods \citep{zhang2024disentangling}. Experiments using ECAPA-TDNN embeddings have found systematic relationships between the age difference separating enrollment and test recordings and the resulting verification scores. The effects also differed across languages and between male and female speakers \citep{singh2023speaker}.

Variation in system performance is not limited to broad recording conditions. It also occurs across individual speakers. A comparison of GMM-UBM, i-vector, x-vector, and ECAPA-TDNN systems found that improvements in average system performance did not apply equally to every person. Some speakers remained difficult to recognize across multiple generations of systems, and both properties of individual recordings and properties associated with particular speakers contributed to performance \citep{harrington2025variability}. Aggregate accuracy can therefore conceal substantial variation in how reliably the same technology represents different individuals.

These results suggest that the same speaker produces a distribution of embeddings rather than a single invariant point. The shape and location of that distribution depend on the speech material, recording conditions, model, and training population. Different speakers may also occupy overlapping regions of the representation space. A speaker embedding is therefore not a voiceprint, despite being frequently labeled or treated as one in the engineering literature \citep{georges2020compact, ronssin2022application, wang2023lightweight, deng2023catch, liu2024real}. It is, instead, one component of a probabilistic comparison system. The appropriate question is not whether two recordings contain the same vocal imprint, but how strongly their observed similarities and differences support competing explanations of their source. This distinction provides the basis for contemporary forensic voice comparison, in which uncertainty, relevant population data, validation, and evidential interpretation must be addressed explicitly.

\section{Speech Deepfakes and Source Speaker Tracing}
\label{sec:deepfake}
\subsection{The Effects of Synthetic Speech on Speaker Identity}
Recent speech synthesis and voice conversion systems can generate novel utterances in a target voice from only a short reference recording \citep{casanova2022yourtts, ma2026meanvc}. These capabilities have already been used to impersonate individuals in a range of contexts, as artificially generated audio becomes increasingly realistic. In 2019, criminals reportedly imitated the voice of a German energy executive and persuaded the chief executive of a UK subsidiary to transfer approximately US\$240,000 \citep{stupp2019fraudsters}. In Hong Kong, an employee transferred approximately HK\$200 million after participating in a fraudulent video conference generated using the impersonated chief financial officer’s voice and video \citep{hongkong2024deepfake}. In Maryland, an AI-generated recording made a high-school principal appear to make racist and antisemitic remarks, leading to his being placed on leave and to police protection at his home because of the threats to him and his family \citep{finley2024deepfake}. Synthetic voices have also been used to attribute political statements to public figures. Before the 2024 New Hampshire presidential primary, voters received robocalls containing an AI-generated imitation of Joe Biden’s voice that discouraged them from voting \citep{fcc2024kramer}. Before Slovakia’s 2023 parliamentary election, a fabricated recording appeared to capture opposition leader Michal Šimečka and a journalist discussing electoral manipulation \citep{unesco2023slovakia}. In each case, the voice was potentially taken as proof that the impersonated person was actually speaking.

Perception studies similarly show that such synthetic recordings cannot be assumed to be easily identifiable. In a study of political speeches, participants identified text-to-speech deepfakes correctly in only about 72\% of observations \citep{groh2024human}. Other listening studies have found similar limitations in human detection. \cite{mai2023warning}, for example, presented genuine and deepfake speech to 529 English- and Mandarin-speaking listeners, who classified the recordings correctly on only 73\% of trials. A study of more than 1,200 participants also reported an average accuracy of 73\% across WaveFake, ASVspoof 2021, and FakeAVCeleb samples \citep{warren2024better}. A meta-analysis of 56 papers involving 86,155 participants estimated an average audio-deepfake detection accuracy of 62.08\%; however, the audio estimate was based on only eight effect sizes and had a wide confidence interval \citep{diel2024human}. Human listeners therefore retain some sensitivity to synthetic speech, but their judgments are not sufficiently consistent to make their perception a reliable indicator of authenticity. Where synthetic generation is a plausible alternative, the interpretation of voice evidence should consider it alongside natural same-speaker and different-speaker explanations.

The perception of speaker identity creates a further problem. Detecting whether a recording is artificial is not the same task as deciding whether it sounds like a particular person. \cite{barrington2025people} tested voice clones from more than 200 speakers and found that listeners judged a clone and its human counterpart to have the same identity on approximately 80\% of trials, but the same listeners correctly identified a recording as AI-generated only about 60\% of the time. Such identity judgments are further affected by linguistic variation, as experiments using cloned voices found that differences in speakers' accents independently biased decisions about whether two recordings were by the same person \citep{yang2026acoustic}. The ability of synthetic speech to convey a target identity is therefore substantial, although it varies across speakers and linguistic conditions.

These findings show that synthetic speech can reproduce enough speaker-related cues for listeners to associate a recording with a particular individual, even when that individual did not produce the utterance. The problem is no longer simply that artificial speech may pass as human; it may pass as the speech of a particular person. That is where the voiceprint metaphor breaks down: a voice can sound convincingly like someone, yet still not have come from that person.

\subsection{Implications for Voice Comparison and Voice Evidence}
Most deepfake-detection systems treat the assessment of suspected deepfake speech as a speaker-independent classification task: a recording is classified as genuine or synthetic without considering whose voice it is intended to reproduce. This framing does not reflect real-world forensic casework involving the suspected impersonation of a known individual. In such a case, the relevant question may instead be whether the questioned recording is a deepfake of that particular speaker. \cite{yang2025forensic} addresses this question within a likelihood-ratio framework, showing that the acoustic differences between genuine and cloned speech are speaker-specific rather than uniform across individuals. By using interpretable phoneme-level features commonly examined in forensic voice comparison, the study also makes the evidential basis of the analysis open to scrutiny rather than relying solely on the output of a generic black-box classifier.

Recent studies support this move from generic classification toward speaker-centred phonetic analysis. A person-of-interest method constructed phoneme-level reference profiles and showed that deviations in synthetic speech could be localized to particular phonemes \citep{salvi2025phoneme}, while a phoneme-aligned dataset found that the most discriminative units varied across phoneme classes and synthesis systems \citep{nallaguntla2026phonemedf}. A more recent explainability study further supported this framing by aligning detector attributions with phoneme boundaries. It found that the cues supporting bona fide and synthetic classifications varied significantly across speakers and phonetic contexts \citep{taylor2026you}. These studies suggest that the acoustic differences between genuine and synthetic speech vary across speakers, phonetic contexts, and generation systems. Such differences may also reflect what a system has learned from its training data rather than an inherent property of synthetic speech \citep{yang2026assessing}. Deepfake detection therefore does not reveal a single acoustic signature of synthetic speech, nor does speaker-specific analysis recover a unique voiceprint.

This has direct consequences for forensic voice comparison. A high degree of similarity between a questioned recording and known recordings may arise because the known speaker produced the utterance, but it may also arise because a generative system reproduced characteristics learned from that speaker \citep{li2024sef, ji2025controlspeech}. The comparison may therefore indicate that the recordings share speaker-related properties without establishing how those properties entered the questioned signal. A close vocal match is not, by itself, evidence that the apparent speaker performed the act of speaking represented in the recording.

In the current era, the interpretation of voice evidence must therefore consider artificial generation alongside natural same-speaker and different-speaker explanations. In suspected impersonation cases, deepfake analysis extends speaker comparison by asking not only whether the questioned voice resembles the known speaker, but also whether that resemblance arose from the speaker's own speech or from an artificial reproduction of the speaker's characteristics. Audio-authenticity analysis may provide additional evidence about editing or processing, but it does not answer the same question. The voiceprint metaphor is particularly misleading in this context because it assumes that a recognizable voice must have been produced by the person it resembles.

\section{Conclusions and Recommendations}

\subsection{Why the Voiceprint Concept Fails Scientifically}

The voiceprint concept fails scientifically because it rests on several assumptions that are inconsistent with the nature of speech production, speaker variation, and modern speaker recognition systems.

First, there is no single stable acoustic object that constitutes personal identity. A voice recording is an observation of a particular speech event produced under particular linguistic, physiological, social, and recording conditions. It is not a direct measurement of a permanent identifying property of an individual.

Second, within-speaker variability is a fundamental property of speech production rather than a minor source of measurement noise. The same speaker does not reproduce an identical acoustic pattern across utterances, situations, or periods of life.

Third, acoustic features rarely have a single cause. They may simultaneously reflect properties of the speaker, linguistic content, speaking style, social context, physical condition, and recording process. Automatic representations introduce further dependencies on training data, model architecture, and evaluation procedures. Speaker-related information therefore cannot be assumed to be completely separable from all other sources of variation or interpreted as a fixed identity imprint.

Fourth, even a high degree of similarity does not establish identity by itself. Similar recordings may originate from different speakers, while recordings from the same speaker may differ substantially. Similarity and difference must instead be interpreted relative to competing explanations, relevant populations, and the conditions under which the evidence has been evaluated.

Fifth, any claim that a recording identifies one person to the exclusion of all others requires a level of empirical validation and evidential support that cannot be inferred from discrimination performance alone. The scientifically defensible conclusion is therefore not that a recording uniquely identifies a person, but that the observed evidence provides a particular degree of support for one proposition relative to another.

The central conclusion of this review is therefore:

\begin{quote}
The terms "voiceprint" and "voiceprint identification" are scientifically misleading because they transform a dynamic, context-sensitive, and probabilistic source of speaker-related information into an imagined, fixed, and unique mark of personal identity.
\end{quote}

\subsection{Terminology and Reporting Recommendations}

Because terminology shapes interpretation, researchers and practitioners should use language that accurately reflects the evidential status and limitations of voice-based conclusions.

The term \textit{voiceprint} or \textit{voiceprint identification} should generally be restricted to historical or critical discussion. Expressions such as \textit{match} or \textit{identification} should be used cautiously when they imply that a recording uniquely identifies one person to the exclusion of all others without sufficient empirical validation.

When referring to the human voice as a source of information about speaker identity, the terms \textit{voice} or \textit{speech} are generally sufficient. More specific expressions such as \textit{vocal properties}, \textit{acoustic features}, \textit{speaker-related information}, or \textit{speaker-discriminative information} may be used when the relevant properties of the speech signal need to be described explicitly. These terms indicate that speech contains information that can contribute to speaker differentiation without implying the existence of a fixed, unique, and permanent vocal imprint.

In automatic systems, learned speaker representations should be interpreted as task-specific representations derived under particular training and evaluation conditions rather than as direct measurements of an intrinsic vocal identity. Reports of speaker-recognition performance should therefore consider relevant sources of variability, including recording conditions, linguistic content, speech duration, speaking style, demographic variation, and domain mismatch.

When recordings are examined to assess whether they may have originated from the same speaker, the process should be described as \textit{voice comparison} or, in a forensic context, \textit{forensic voice comparison}. In automatic systems, \textit{speaker verification} refers to assessing a claimed speaker identity, while \textit{speaker identification} refers to selecting a possible speaker from a defined set of candidates. \textit{Speaker recognition} may be used as the broader term covering both tasks. These terms more accurately describe the underlying processes and avoid misleading analogies with fingerprints.

Reports should clearly state the competing propositions, the relevant population, the methods used, and the conditions under which those methods were validated. They should describe important limitations arising from recording quality, channel mismatch, speech duration, linguistic material, speaking style, and possible manipulation. Where numerical likelihood ratios or probabilistic scores are reported, both discrimination and calibration performance should be documented under relevant conditions. Speaker-source evaluation should also be clearly distinguished from audio-authenticity analysis when necessary, since similarity to a known speaker does not by itself establish that the speaker produced the questioned utterance.

\begin{comment}
\section*{Acknowledgements}
We thank Matthew Faytak for his helpful comments and suggestions on the conceptualization and presentation of the speech-chain model shown in Figure~1.

\section*{Funding}
Cuiling Zhang was supported by the Natural Science Foundation of Chongqing, China (CSTB2022NSCQ-LZX0007), and the Key Project of the Science and Technology Research Program of the Chongqing Municipal Education Commission (KJZD-K202600302).
\end{comment}

\bibliographystyle{elsarticle-harv}
\bibliography{references}

\end{document}